\documentclass[12pt]{article}
\usepackage[pdftex]{graphicx}
\usepackage{color}
\usepackage{amsmath}
\begin{document}
\newsavebox{\savepar}
\newenvironment{boxit}{\begin{lrbox}{\savepar}
\begin{minipage}[b]{7in}}
{\end{minipage}\end{lrbox}\fbox{\usebox{\savepar}}}
\newcommand{\be}{\begin{equation}}
\newcommand{\ee}{\end{equation}}
\newcommand{\red}{\color{red}}
\newcommand{\blue}{\color{blue}}
\newcommand{\black}{\color{black}}
\parskip 10pt
\parindent 20pt

\font\script=eusm10
\font\bbold=msbm10
\input amssym.def
\input amssym

\newcommand{\SW}{\mbox{\script W}}
\newcommand{\SC}{\mbox{\script C}}
\newcommand{\ScS}{\mbox{\script S}}
\newcommand{\ScL}{\mbox{\script L}}
\newcommand{\SO}{\mbox{\script O}}
\newcommand{\SP}{\mbox{\script P}}
\newcommand{\BM}{\mbox{\bbold M}}
\newcommand{\BL}{\mbox{\bbold L}}
\newcommand{\SM}{\mbox{\script M}}
\newcommand{\BR}{\mbox{\bbold R}}
\newcommand{\BC}{\mbox{\bbold C}}

\title{Principal Component Analysis for Equation Discovery}

\vspace{-1.0cm}

\author{Caren Marzban$^{1,2}$\thanks{Corresponding Author: marzban@stat.washington.edu}, 
Ulvi Yurtsever$^3$, Michael Richman$^4$ \\ \\
$^1$ Applied Physics Laboratory, $^2$ Department of Statistics \\
Univ. of Washington, Seattle, WA 98195 USA \\
$^3$ MathSense Analytics, 1273 Sunny Oaks Circle, Altadena, CA, 91001, USA\\
$^4$ School of Meteorology, University of Oklahoma, Norman, OK 73072, USA
}

\date{ }
\maketitle
\vspace{0.0cm}

\begin{abstract}
\parindent 0pt 

\end{abstract}
Principal Component Analysis (PCA) is one of the most commonly used statistical methods for data 
exploration, and for dimensionality reduction wherein the first few principal components account 
for an appreciable proportion of the variability in the data. Less commonly, attention is paid to 
the last principal components because they do not account for an appreciable proportion of 
variability. However, this defining characteristic of the last principal components also qualifies
them as combinations of variables that are constant across the cases. Such constant-combinations are 
important because they may reflect underlying laws of nature.  In situations involving a
large number of noisy covariates, the underlying law may not correspond to the last principal 
component, but rather to one of the last. Consequently, a criterion is required to identify the
relevant eigenvector. In this paper, two examples are employed to demonstrate the proposed methodology; 
one from Physics, involving a small number of covariates, and another from Meteorology wherein the 
number of covariates is in the thousands. It is shown that with an appropriate selection criterion,
PCA can be employed to ``discover" Kepler's third law (in the former), and the hypsometric equation
(in the latter).


\parindent 20pt
\newpage
\section{Introduction}

Given data on a number of variables, it is often desirable to determine whether some of the 
variables are related to one another in a useful manner. The use may be for dimensionality reduction,
prediction purposes, or for discovering a law of nature. Inferring and understanding structure in 
large data sets has been an active area of research. Although it has been called by various names, 
Knowledge Discovery in Data bases, Data Mining, Deep Learning, and Equation Discovery are some of 
common names (Bergen, et al. 2019; Bongard and Lipson 2007; Grundner et al. 2023; 
Hoerl, Snee, and De Veaux 2014; Kamath 2001; Kratzert et al. 2019; Kurgan and Musilek 2006; 
Langley 1981; Marzban and Yurtsever 2011; 2017; Schmidt and Lipson 2009; Song et al. 2023; Wang 1999; 
Xu and Stalzer 2019; Yu and Ma 2021; Zanna and Bolton 2020; Zhang and Lin 2018). A subset of 
these works aims to discover - from data - physical laws of nature in terms of relatively
simple algebraic equations. For example, the work of Bongard and Lipson (2007), 
Schmidt and Lipson (2009), and Xu and Stalzer (2019) is based on a symbolic regression approach 
wherein a space of compact algebraic expressions is searched for an expression that fits the data. 
One of the most recent works explores automatic determination of the number of state variables, and 
what they may be, directly from video streams (Chen et al. 2022). The underlying methods are wide-ranging, but Camps-Valls et al. (2023) provides a summary and taxonomy. 

The equation discovery methods consider a wide range of equation types, including linear, nonlinear, 
differential, and partial differential. Although a wide range of scientific fields is examined in 
these works, Meteorology is not prominently represented among them. Consequently, one purpose of the
present article is to consider a demonstration of equation discovery that is familiar to 
meteorologists. That said, the aim is not to produce a review article, but rather to propose yet 
another equation discovery method, specifically one that is based on a technique commonly
employed in meteorology, namely Principal Component Analysis (PCA), often referred to as Empirical
Orthogonal Functions. 

One key observation underlying the proposed approach is that a useful relationship between variables 
will manifest itself as some function of the variables that is constant across the cases in the data. 
For example, in a data set involving the gravitational force ($F$) between two objects with masses $M$ 
and $m$, separated by a distance $r$, Newton's law of universal gravitation, $F = G M m/r^2$, would be 
manifested as $Fr^2/(Mm) = G$, where $G$ is the gravitational constant.  In a practical setting, the 
data on the variables $F, M, m, r$ are subject to ``noise," as a result of which the specific 
combination $Fr^2/(Mm)$ will not be exactly equal to the value of $G$. In passing, note that the
nonlinear equation $Fr^2/(Mm)=G$ can be rendered linear by examining the logarithm of the variables.

The question of whether a certain combination of variables is useful is central to the method of 
PCA (Hotelling 1933; Jolliffe 2002; Lorenz 1956; Wilks 2019). The most common application of PCA is to 
identify the most-varying combinations of variables. Consequently, PCA is most often employed 
for dimensionality reduction, because the combinations with the largest variability encapsulate 
the information in the data with fewer variables (at least, if the number of variables is less
than the number of cases - a condition assumed throughout this work). PCA in its most common form
leads to a set of variable combinations - called Principal Components (PC) - equal in number to the 
original set of variables, that can be rank-ordered in terms of the amount of the variability 
explained by each. As such, a subset of PCs consisting of the highest-ranking PCs summarizes the data 
with fewer variables. The variance explained by each PC, and the coefficients of variables in 
each PC - often called loadings - are given by the eigenvalues and eigenvectors, respectively, of the 
sample covariance matrix. In situations where the variables have significantly different variances, 
it is common practice to examine the eigendecomposition of the correlation matrix. 

The second key observation for the proposed method is as follows: 
Given that equation discovery involves finding combinations of variables that are constant, 
PCA is appropriate because the lowest-ranking PCs are combinations with the lowest variability.
Situations in which the focus of PCA is on the lowest-ranking combinations have been described,
and are often referred to as ``Last PCA," or "Least PCA" (Gertler and Cao 2005; Huang 2001; 
Jolliffe 2002; Jolliffe an Cadima 2016; Rolle 2002; Zlobina and Zhurbin 2020). Although in one of 
the examples considered here,
the relevant equation does in fact appear in the last eigenvector, in general the existence of
sampling variability (i.e., noise) implies that the equation of interest may be associated
not with the last eigenvector but with one of the last eigenvectors. Indeed, some number of the last 
eigenvectors are often degenerate, i.e., have equal eigenvalues (Anderson 1963; North et al. 1982). 
Moreover, it is possible that the data encapsulate multiple laws, or no laws, at all.
Although there exist numerous methods for assessing how many of the highest-ranking PCs ought to be 
saved (Ibebuchi and Richman 2023; Wilks 2016), not only such methods do not exist for 
lowest-ranking PCs, but more importantly, the identification of relevant PCs requires consideration 
of physical significance in addition to statistical significance.  Therefore, the identification of
the relevant eigenvector(s) calls for some sort of a selection criterion, which will be discussed here.
With that understanding, henceforth, the ``last" eigenvector refers to one of the last eigenvectors.

In short, therefore, an approach to the discovery of equations entails performing PCA on data, 
and identifying the last PC. That specific combination is, then, the most constant (i.e., least 
variable) combination of the variables. And the numerical value of that constant can be estimated by 
the sample mean of the data projected onto the last PC (often called PC score). Another advantage of 
employing PCA for equation discovery is that PCA treats all variables on the same footing, unlike 
regression, where the variables must be divided into two sets - predictors and response(s).

Here, such a PCA-based equation discovery method is presented. The outline of the paper is as follows: 
The Method section provides a high-level presentation of the proposed methodology; this section also
includes a subsection wherein a formula is derived relating PCA loadings to parameters of the 
underlying equation; details of this derivation are contained in the Appendix. Two examples 
are then considered where further details of the proposed method are revealed. The second example 
proposes a selection criterion for the last PC, appropriate for situations where the variables are 
spatial fields - common in Meteorology. The paper ends with a summary of the conclusions, and a 
discussion of the limitations and generalizations of the method.

\section{Method}

Given data on $p$ variables, there are several ways of discovering underlying relationships.
At the simplest level, one can simply examine the correlation coefficient between all pairs of
variables. If, however, one is interested in a relationship that involves multiple variables, then at 
the broadest level, the methods of choice are regression and PCA. For the specific purpose of equation
discovery, the main disadvantage of the former is that it requires the specification of a response
variable, or a set of response variables, which {\it a priori} may be unknown. That said, the main
advantage of regression is that the parametric form of regression models naturally aligns with many
laws of nature. For example, fitting a regression model of the type $y = \alpha + \beta x + \epsilon$ 
to data on acceleration $x$, and force $y$, will readily ``discover" $F=ma$, with the regression 
coefficient $\beta$ estimating mass $m$. Alternatively, one may fit the regression model 
$y = \alpha + \beta x + \epsilon$ to
data on the logarithm of acceleration and force, in which case the parameter $\alpha$ would 
estimate the logarithm of $m$, and the slope parameter $\beta$ would be approximately 1.
This natural alignment of the regression line and physical laws is a consequence of the fact
that the regression line is designed to estimate the conditional mean of $y$, given $x$.

By contrast, the parameters in the PCA line $y = a + b x$ cannot be readily identified with
physical parameters, because the PCA line is not designed to estimate the conditional mean
of $y$, given $x$. Instead, it is designed such that its slope points in the direction of
maximum variability.\footnote{It can be shown that the PCA line minimizes the sum-square of the
{\bf shortest} distances between the data and the line.} However, the main advantage of PCA is that
it treats all variables on the same footing, without requiring the separation of variables
into responses and predictors. As shown in the next section, it is possible to find analytic, 
closed-form expressions that relate the parameters of the PCA line to those of the regression
line. Therefore, for the purpose of equation discovery it is more natural to use PCA first, and then
use the analytic formulas of the next section to infer the parameters of the regression line,
i.e., the parameters that naturally align with physical parameters.

\subsection{From PCA to Regression}

Most often, PCA involves an eigendecomposition of either the covariance or the correlation matrix
(although exceptions do exist, e.g., Elmore and Richman 2001). For demonstration purposes consider
a data set involving only two variables, $x$ and $y$. Although the assumption of normality is
required only for assessing the statistical significance of the results, for convenience assume
that the data follow a bivariate normal distribution with mean parameters $\mu_x, \mu_y$, 
variance parameters $\sigma_x^2, \sigma_y^2$, and correlation parameter $\rho$. The covariance 
matrix can then be written as
\begin{equation}
\Sigma = \left( \begin{array}{cc}
 \sigma_x^2 & \rho \; \sigma_x \sigma_y  \\
 \rho \; \sigma_x \sigma_y  & \sigma_y^2  \end{array} \right) .
\end{equation}
The two eigenvectors of $\Sigma$ can be found to be
\begin{equation}
\left( \begin{array}{cc}
 1   & 1  \\
 A_{+} & A_{-}  \end{array} \right),
\end{equation}
where $A_{\pm} =  A \pm \sqrt{1+A^2}$, and 
$A = \frac{1}{2\rho} ( \frac{\sigma_x}{\sigma_y} - \frac{\sigma_y}{\sigma_x} )$. In practice,
all of these distribution parameters are estimated by their sample analogs. These two eigenvectors
define the direction (slope) of two PCA lines. The eigenvector with the larger eigenvalue points
in the direction of maximum variability, and the second eigenvector is orthogonal to the first one.
The y-intercept of the lines is fixed by the criterion that the line must go through the point
$(\mu_x, \mu_y)$. In short, the equations of the PCA lines are $y = a + b x$, with 
$b = A_{\pm}$. A formula for the y-intercept is also available, but is not given here because 
the slope parameter is the only parameter of interest.

Figure 1 shows an instance of data (circles) from a bivariate normal with parameters 
 $\mu_x = 0, \mu_y = 10, \sigma_x^2 = 2, \sigma_y^2 = 3, \rho = 0.8$. The two red lines denote
the PCA lines. For comparison, the regression line is also shown (in green).

As mentioned previously, the slope that is more relevant to equation discovery is that of the regression 
line. Given that the slope of the regression line is $\beta = \rho \frac{\sigma_y}{\sigma_x}$, it is 
easy to show that 
\begin{equation}
\beta = \frac{1}{2} [ -B \rho^2 \pm \sqrt{ B^2 \rho^4 + 4 \rho^4} ] \;,
\end{equation}
where $B=(b-\frac{1}{b})$, and $b$ denotes the slope of the PCA lines, i.e., $A_{\pm}$.

\centerline{ \includegraphics[height=3in,width=3in]{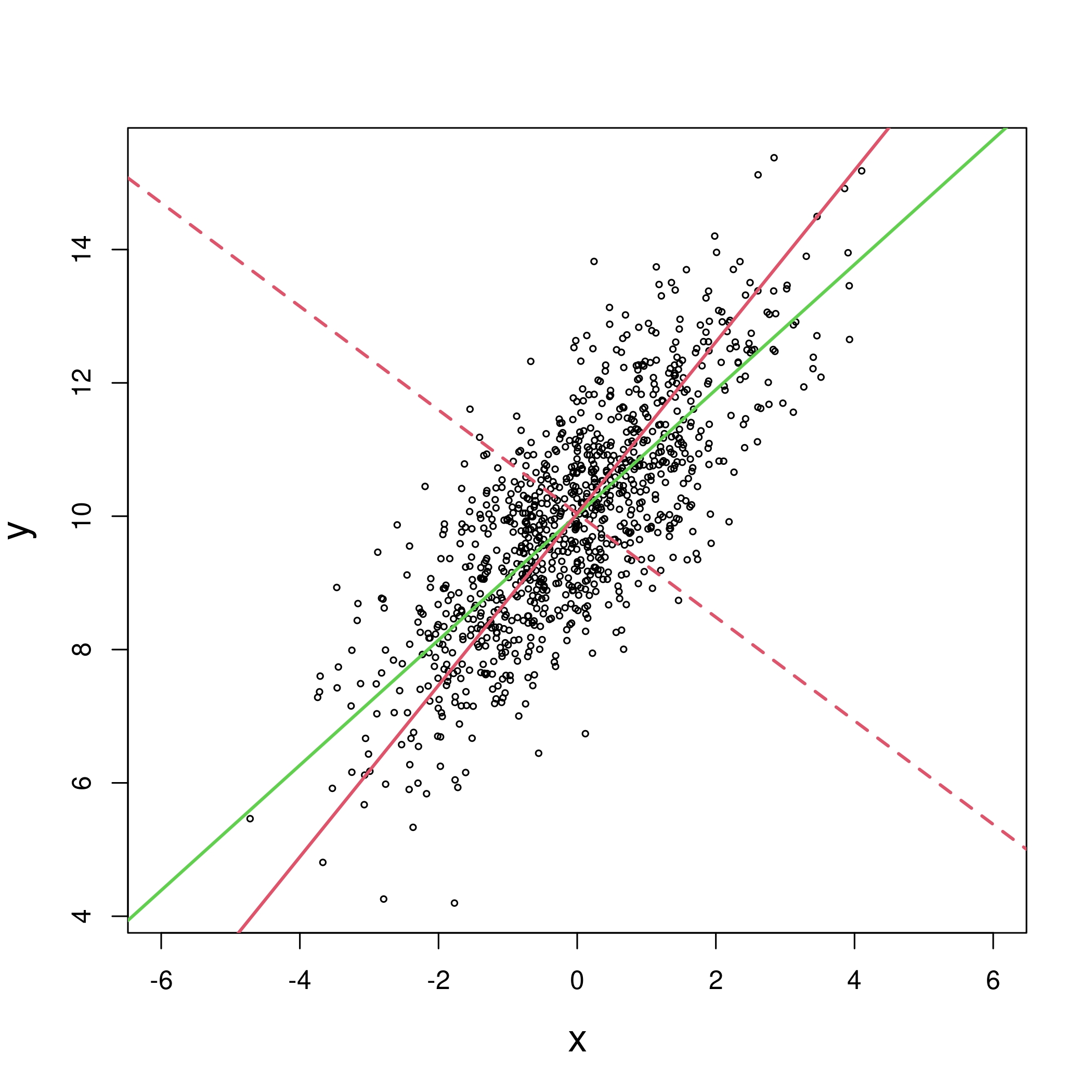} }
Figure 1. A demonstration of the PCA lines (in red), and the regression line (in green).

As seen in Figure 1, for sufficiently strong correlation between the variables, the PCA line 
corresponding to the largest eigenvalue is close to the regression line, and the PCA line with the
smallest eigenvalue is orthogonal to the regression line. Therefore, in such situations the 
loadings can be used directly to infer an underlying law, without use of Eq. 3. Example 1 (below)
demonstrates this scenario. However, more generally (as in Example 2), Eq. 3 will be used to 
estimate the parameters of the underlying law. 

\section{Examples}

To demonstrate the above methodology for equation discovery, two examples are considered - one
from Physics and Astronomy, and one from Atmospheric Sciences. The rationale for the specific choice 
of the examples is as follows: First, to illustrate the generality of the methodology, the examples 
are taken from different fields.  Second, Example 1 involves a small number of variables and a small 
sample size - 5 and 8, respectively. The goal is to show that the last eigenvector does in fact 
correspond to a known physical law, namely Kepler's third law.  By contrast, the second example 
involves several hundred covariates because three variables are observed at hundreds of grid
points across space. In such a situation
the last eigenvector may not play an important role at all, because it is possible to construct
linear combinations that have even less variability than the combination corresponding to the law.
Therefore, it is necessary to introduce a criterion for identifying the specific eigenvector 
corresponding to the underlying law, which in the second example is the hypsometric equation.

The task of identifying one of the last eigenvectors that represents a law is not dissimilar to
that of selecting some number of the largest eigenvalues that adequately summarize the data with
fewer variables. The latter has been studied extensively (Jolliffe 2002), and has given 
rise to a number of commonly used criteria. One criterion is to select a sufficiently large number 
of the eigenvectors with the largest eigenvalues that explain some specified percentage of the total 
variability.  Another criterion is to select the eigenvectors up to that corresponding to the ``elbow" 
in the scree plot (Dmitrienko, Chuang-Stein, and D'Agostino 2007). Other criteria are based on
statistical significance (Cattell 1966; Wilks 2016). Furthemore, Ibebuchi and Richman (2023) point
out that none of these criteria are based on physical considerations of whether a given PC ought
to be kept. In short, there does not exist a universal 
criterion that leads to a unique number of eigenvectors that adequately summarize the data - a 
situation arising in the present application, as well. Here, a criterion for selecting the 
physically relevant eigenvector is introduced in Example 2.

\subsection{Example 1}

Consider Kepler's third law of planetary motion $\frac{G(M+m)T^2}{a^3} = 4\pi^2$, 
where $G$ denotes Newton's gravitational constant, $M$ and $m$ denote the masses of the Sun
and of the orbiting planet, and $T$ and $a, \; b$ denote the period 
and the semi-major and semi-minor axes of the planet's elliptical orbit, respectively.
If this law were to be discovered 
empirically, the data would involve 8 observations (for 8 planets) of the variables $a, b, m, M, T$. 
The semi-minor axis $b$ has been included, allowing for the possibility that 
the underlying relationship involves both the semi-major and semi-minor axes. Note that for the solar
system $M$ --- the central solar mass for all planetary orbits --- is a constant,
$M = 1.986616 \times 10^{30} (Kg)$, and $M \gg m$ 
for any of the planets (for Jupiter $m/M \approx 10^{-3}$). Therefore, in this work
Kepler's law can be taken to be in the form $\frac{G M T^2}{a^3} = 4\pi^2$.
The data are shown in Table 1.

Given that PCA involves linear combinations, it is necessary to perform PCA on the logarithm of 
data. In this log-space, Kepler's law becomes $2 \log{T} - 3 \log{a} = $ constant.  Performing PCA 
on the data shown in Table 1 yields the eigenmatrix shown on the left in Eq.\,(4), with the columns
denoting the eigenvectors, and the rows corresponding to the variables $a, b, m, M,$ and $T$, 
respectively. 
These eigenvectors are orthonormal, but they can be ``normalized" differently (as discussed in the 
discussion section) to expose the underlying near-integer values (right side of Eq.\,4). For example, 
it follows that the first PC involves $a^{2.00} b^{2.01} m^{3.83} M^{0.00} T^{3.00}$.

\begin{table}
\begin{center}
\caption{Data for the solar system. }
\begin{tabular}{lllll}
Planet  & $a (\times 10^{10} m)$ & $b (\times 10^{10} m)$ & $m (10^{24} Kg)$ & $T (Sec)$ \\ \hline
Mercury & 5.852857 & 5.727818 & 0.3244425 & 7605382 \\
Venus   & 10.81012 & 10.80988 & 4.861260  & 19407924 \\
Earth   & 14.95104 & 14.94896 & 5.975000  & 31557600 \\
Mars    & 22.82995 & 22.73016 & 0.6387275 & 59359846 \\
Jupiter & 77.82562 & 77.73441 & 1902.141  & 374336251 \\
Saturn  & 142.7208 & 142.4993 & 569.4175  & 929623781 \\
Uranus  & 287.0700 & 286.7501 & 87.11550  & 2651311764 \\
Neptune & 449.5683 & 449.5517 & 103.1285  & 5200313789 \\
\end{tabular} 
\end{center}
\end{table}

\begin{equation}
\left(
\begin{array}{rrrrr}
 -0.36 & 0.33 & 0.72 & -0.50 & 0 \\
 -0.36 & 0.33 &-0.70 & -0.53 & 0 \\
 -0.68 &-0.73 & 0.00 &  0.00 & 0 \\
  0.00 & 0.00 & 0.00 &  0.00 & 1 \\
 -0.53 & 0.49 &-0.01 &  0.69 & 0 
\end{array}
\right) 
\sim
\left(
\begin{array}{ccccc}
 2.00 &  2.00 &  1.00 &  3.00 & 0 \\
 2.01 &  2.00 & -0.97 &  3.15 & 0 \\
 3.83 & -4.45 &  0.00 &  0.00 & 0 \\
 0.00 &  0.00 &  0.00 &  0.00 & 1 \\
 3.00 &  3.00 & -0.02 & -4.10 & 0
\end{array} 
\right)
\end{equation}

\begin{table}
\caption{The PCs and the corresponding eigenvalues of PCA performed on the variables $a, b, m, M, T$.}
\begin{center}
\begin{tabular}{lll} 
$ $ & PC               & eigenvalue\\ \hline
1 & $(ab)^2 T^3 m^4$   & 18.42139 \\
2 & $(ab)^2 T^3 / m^4$ & 2.509082 \\
3 & $(a/b)^3$          & 0 \\
4 & $(ab)^3 / T^4$     & 0 \\
5 & $M$                & 0
\end{tabular}
\end{center}
\end{table} 

Upon rounding the loadings to the nearest integer, the resulting PC and the corresponding eigenvalues 
are shown in Table 2.  Evidently, the first eigenvector in Eq. (4) suggests that the combination 
$(ab)^2 m^4 T^3$ has the largest variability. The second eigenvector implies that the second 
most-variable combination is $(ab)^2 T^3 / m^4$. According to the next eigenvector, which has zero 
eigenvalue, $(a/b)$ is a 
constant, a consequence of the near-circular orbits in the solar system. The last two 
eigenvectors, also with zero eigenvalue, imply that $(ab)^3 / T^4$ and $M$ are constants. The latter 
is a consequence of the aforementioned fact that the planets all orbit the same sun in the solar system. 
Kepler's law is, therefore, embodied in the penultimate eigenvector. The sample mean of the 
corresponding principal score is $87.45$, and since $a \sim b$, it follows that 
$\log(a^3/T^2)^2 = 87.45,$ i.e., $\log(a^3/T^2) = 87.45/2= 43.75$. This value is within 
observational error of $\log(GM/(4\pi^2)) = 42.66$, as expected from Kepler's law. 

As a test of the assumptions of PCA, the scatterplots of the variables $(a, b, m, M, T)$ (after log) 
are examined. It is found that all of the variables are linearly related, with the exception of $M$, 
of course.  The pairwise correlation coefficients vary between 0.76 (between $a$ and $m$) and 1.00 
(between $a$ and $b$). As such, there is no evidence that the linearity assumption is violated.

\subsection{Example 2}

PCA is often performed on 2-dimensional fields. In image processing circles, the fields are images,
i.e., a collection of pixels; in atmospheric sciences PCA is often performed on gridded fields, where 
the variables are ``observed" at each of several thousand grid points across a 2-dimensional spatial 
lattice. In such applications, the main purpose of PCA is to identify the dominant spatial 
structures, often called eigenfaces in image processing.  Equation discovery, then, requires performing
PCA on multiple spatial fields, each corresponding to a different physical variable. For example, two 
physical variables may be surface temperature and surface pressure. Here, three physical variables
are examined: 1) thickness (the vertical distance between two pressure levels), denoted $H$, 
2) mean virtual temperature (the temperature at which a theoretical dry air parcel would have a 
total pressure and density equal to the moist parcel of air), denoted $T_v$, and 3) meridional wind 
speed, denoted $V$. The first two physical variables are in fact known 
to follow a ``law" known as the hypsometric equation (Wallace and Hobbs 1977): 
\begin{equation}
H = (R/g) log(p_1/p_2) T_v \;,
\end{equation}
where $R = 2.87 * 10^2 (m^2/(s^2 K))$ is the specific gas constant for dry air, $g = 9.8 (m/s^2)$ is
acceleration due to gravity, and $p_1, p_2$ are the pressures at the two levels whose thickness
is $H$. Here, the two pressure levels are selected to be $850 (HPa)$ and $500 (HPa)$.

The goal of this example is to demonstrate how PCA can be employed to ``discover" the hypsometric
equation (dictating $H$ and $T_v$) from gridded data on $H, T_v$, and $V$. The variable $V$ is
included in the analysis because in law discovery one does not know what variables are dictated 
by the law {\it a priori}. To that end, data are obtained from the Reanalysis project
(Kalnay et al. 1996) where all fields are given on a $144 \times 73$ grid covering the entire globe. 
As for the temporal scale of data, focus is placed on the 530 monthly means between Jan 1, 1979 and 
March 1, 2023.  The virtual temperature is not provided in the Reanalysis data base, and therefore, it 
is computed from the approximate formula (Doswell and Rasmussen 1994; Glickman 2000)
$T_v  = T \frac{1 + q/0.622}{1 + q}$ where $T$ and $q$ are the 
temperature and specific humidity at a given pressure level, both available in the Reanalysis database. 
To assure the underlying relationships between the variables are not overwhelmed by the
periodic nature of the data, the 12-month period is filtered out using a difference filter, leading
to $530 - 12 = 518$ cases.
Figure 2 shows the mean (across all 518 months) of the resulting residual fields.

\centerline{ \includegraphics[height=5in,width=5in]{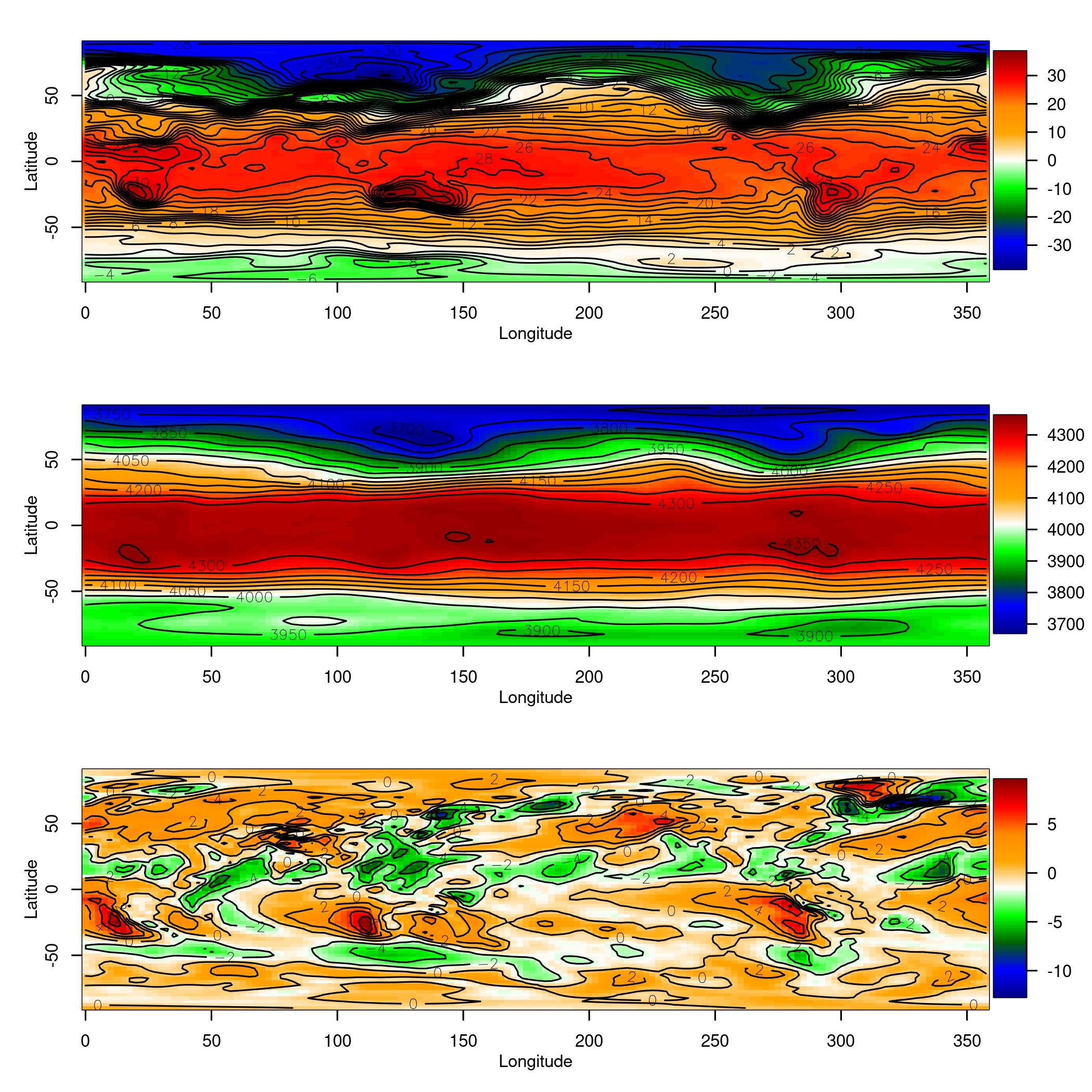} }
Figure 2. From top to bottom, the mean (across 518 months) of the monthly mean gridded data for 
virtual temperature (in Centigrade), thickness (in meters), and meridional wind speed (in $m/s$),
after the 12-month period is filtered out.

Performing (S-mode) PCA on a temporal sequence of gridded fields requires ``flattening" each
image into a vector of inputs (or feature vectors). For example, if PCA is performed on the gridded 
field of thickness, with 144 grid points along the longitude and 73 grid points along the latitude, 
then the dimensionality of the input vector is $144 \times 73 =$ 10,512. 
Performing PCA on all three fields combined multiplies the size of the input vector by a factor of 3. 
Although that number is not prohibitive for modern computing, here each field is cropped to only the
northern latitudes from $37.5^{\circ}$ to $77.5^{\circ}$. The resulting spatial domain excludes
the polar and equatorial regions. Given the $2.5^{\circ}$ resolution of the Reanalysis data, this 
leads to a spatial field with dimensions $144 \times 17$ for each of the three physical fields. 
In short, the dimensionality of the input vector for the combined PCA is $3 \times 144 \times 17 =$ 
7,344.  The number of cases in the training data is 518, i.e., the number of months in the database.
Said differently, the dimension of the data matrix is 518 $\times$ 7,344, and the correlation
matrix has dimensions 7,344 $\times$ 7,344, leading to $518-1$ non-zero eigenvalues.\footnote{It can 
be shown that when PCA is performed on $p$ variables and $n$ cases, the number of non-zero eigenvalues 
is the minimum$(p,n-1)$.} 

The results of the combined PCA are as follows: The screeplot, showing the eigenvalues in decreasing 
order, is shown in Figure 3; the square root of the eigenvalues is shown on the y-axis, because that 
quantity measures the standard deviation explained by the corresponding eigenvector.
The two curves correspond to the eigenvalues of the covariance matrix (in black) and those of the 
correlation matrix (in red).  Although the results for only the first 200 eigenvectors are shown, the 
eigenvalues approach zero as the number of eigenvectors approaches 517. The large, filled circles in 
this figure are discussed in the next subsection.

\centerline{ \includegraphics[height=3in,width=3in]{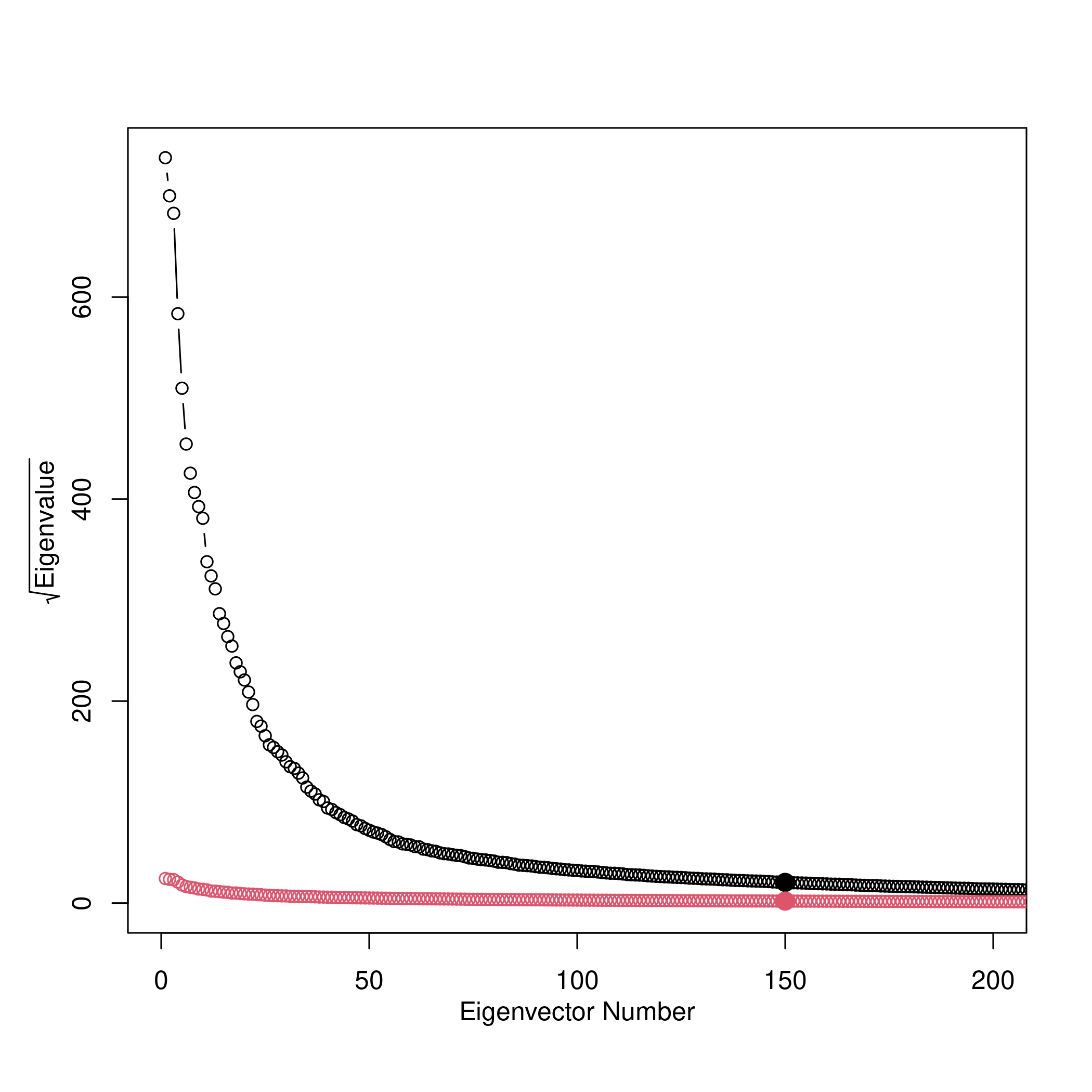} }
Figure 3. The standard deviation explained by the eigenvectors of the covariance (black) and 
correlation (red) matrix. The larger, filled circles are addressed in the subsection Eigenvector 
Selection.

As mentioned previously, in traditional PCA an important question is the smallest number of 
eigenvectors that adequately explain the majority of the variance. As such, the region of interest is 
the left side of the screeplot. In the context of equation discovery, however, the region of interest 
is the right side of the screeplot, because eigenvectors with near-zero eigenvalues define linear 
combinations of variables that are constant across the cases in the data. The question, then, is 
how to identify the eigenvector of interest among all of those with near-zero eigenvalues.

\subsubsection{Eigenvector Selection}

To identify the eigenvector that represents a law, consider the structure 
of the eigenmatrix, shown in Eq. 6. There are 517 columns corresponding to the eigenvectors with
non-zero eigenvalues. There are a total of $3 \times 144 \times 17$ rows corresponding to the
three physical fields, with each field consisting of a spatial grid with 144 grid points along
the longitude and 17 grid points along the latitude. The goal is to identify the column that 
represents the hypsometric equation. Also, to assure that none of the three fields 
dominates an eigenvector simply due to its large variability, henceforth, all eigenvectors are
those of the correlation (not covariance) matrix.

\begin{equation}
 \left( \begin{array}{c|ccccc}
 & e_1 & e_2 & e_3 & \cdots & e_{517} \\ \hline
 & . & . & . & \cdots & .\\
 & . & . & . & \cdots & .\\
T_v & . & . & . & \cdots & .\\
 & \vdots & & & & \\
 & . & . & . & \cdots & . \\ \hline
 & . & . & . & \cdots & .\\
 & . & . & . & \cdots & .\\
H & . & . & . & \cdots & .\\
 & \vdots & & & & \\
 & . & . & . & \cdots & . \\ \hline
 & . & . & . & \cdots & .\\
 & . & . & . & \cdots & .\\
V & . & . & . & \cdots & . \\
 & \vdots & & & & \\
 & . & . & . & \cdots & .
 \end{array} \right) .
\end{equation}

Consider the first eigenvector $e_1$. Note that each of three segments corresponding to three 
physical fields can be displayed as an image (an eigenface) with the same dimensions 
as the original field (Figure 4). In traditional PCA, the spatial structure in these three eigenfaces 
is employed to identify the dominant oscillations. For example, if PCA were done on surface 
temperature, then the first column would be an eigenface related to ENSO. 
Given that PCA is performed on all three variables simultaneously, the correlations between
the physical fields is also taken into account. Said differently, if the three fields were totally
independent of one another, then the three eigenfaces would be exactly the eigenfaces that one
would obtain from PCA performed on the three fields, separately. By the same token, any correlations
between the fields will lead to similar eigenfaces. In short, the similarity of the three eigenfaces
in Figure 4, is a measure of the association between the underlying physical variables. Similarly, 
for the other eigenfaces with smaller eigenvalues.

\centerline{ \includegraphics[height=4in,width=4in]{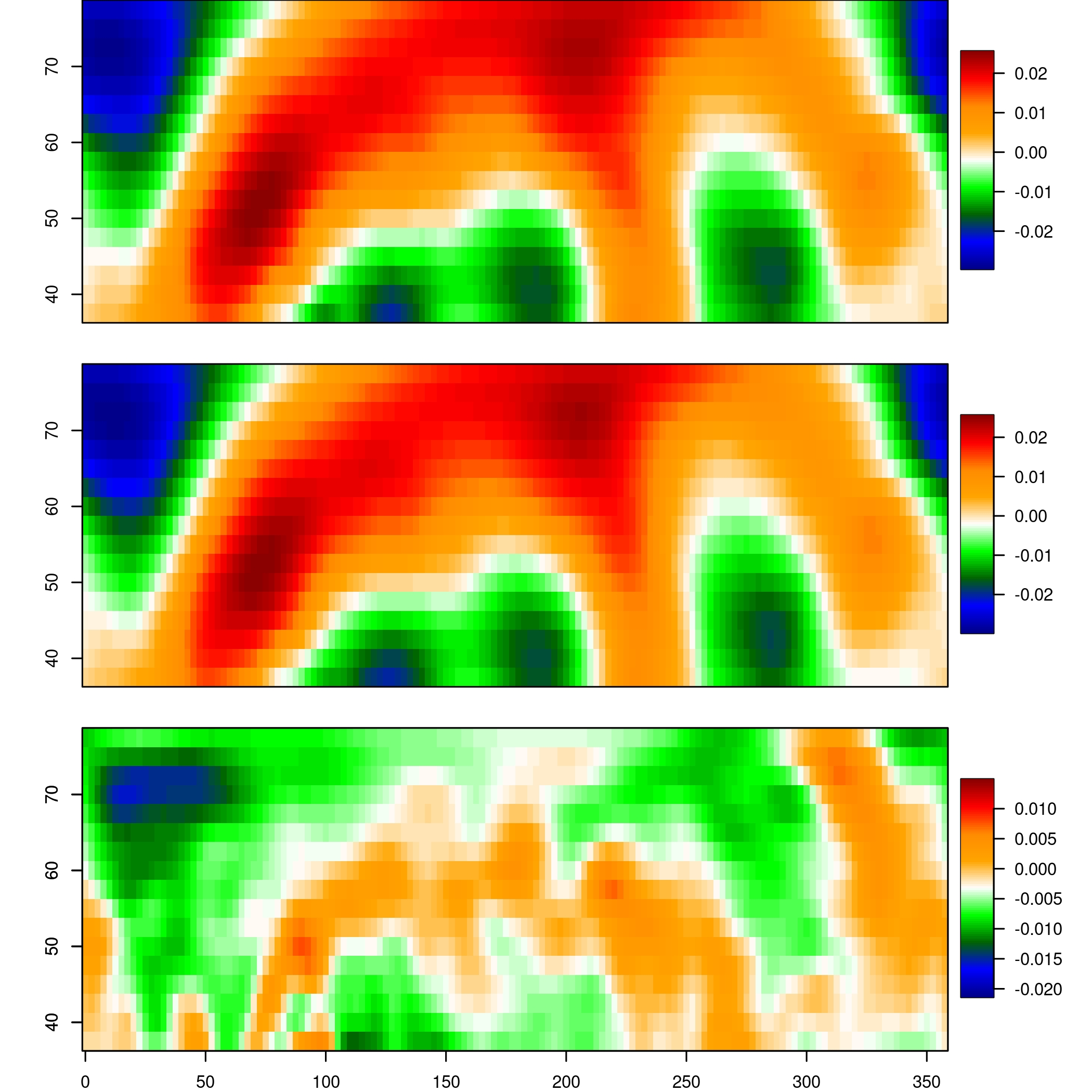} }
Figure 4. The first eigenface from combined PCA on $T_v, H$ and $V$.

Although the spatial structure displayed in eigenfaces is important from the perspective of
understanding the underlying physical processes, for the present application it is necessary and
sufficient to summarize each of the eigenfaces in a way that makes their comparison easier.
To that end, Figure 5 displays the loadings of $e_1$, i.e., the first 
eigenvector ``prior" to display as the eigenfaces in Figure 4. The numbers on the x-axis are the
row number of the eigenmatrix in Eq. 6; the three distinct segments correspond to the three physical 
fields, and the undulations within each segment are a consequence of the spatial structure
in each of the eigenfaces in Figure 4. The advantage of displaying the first eigenface
in this fashion is that one can immediately conclude that the first eigenvector is dominated
mostly by $T_v$ and $H$, with the wind variables $V$ having the smallest loadings. This observation
can be quantified by the variance of the loadings for each of the three fields, separately. 
For example, in Figure 5 it is evident that the right-most segment (corresponding to $V$) has
smaller variance across the loadings.

\centerline{ \includegraphics[height=4in,width=4in]{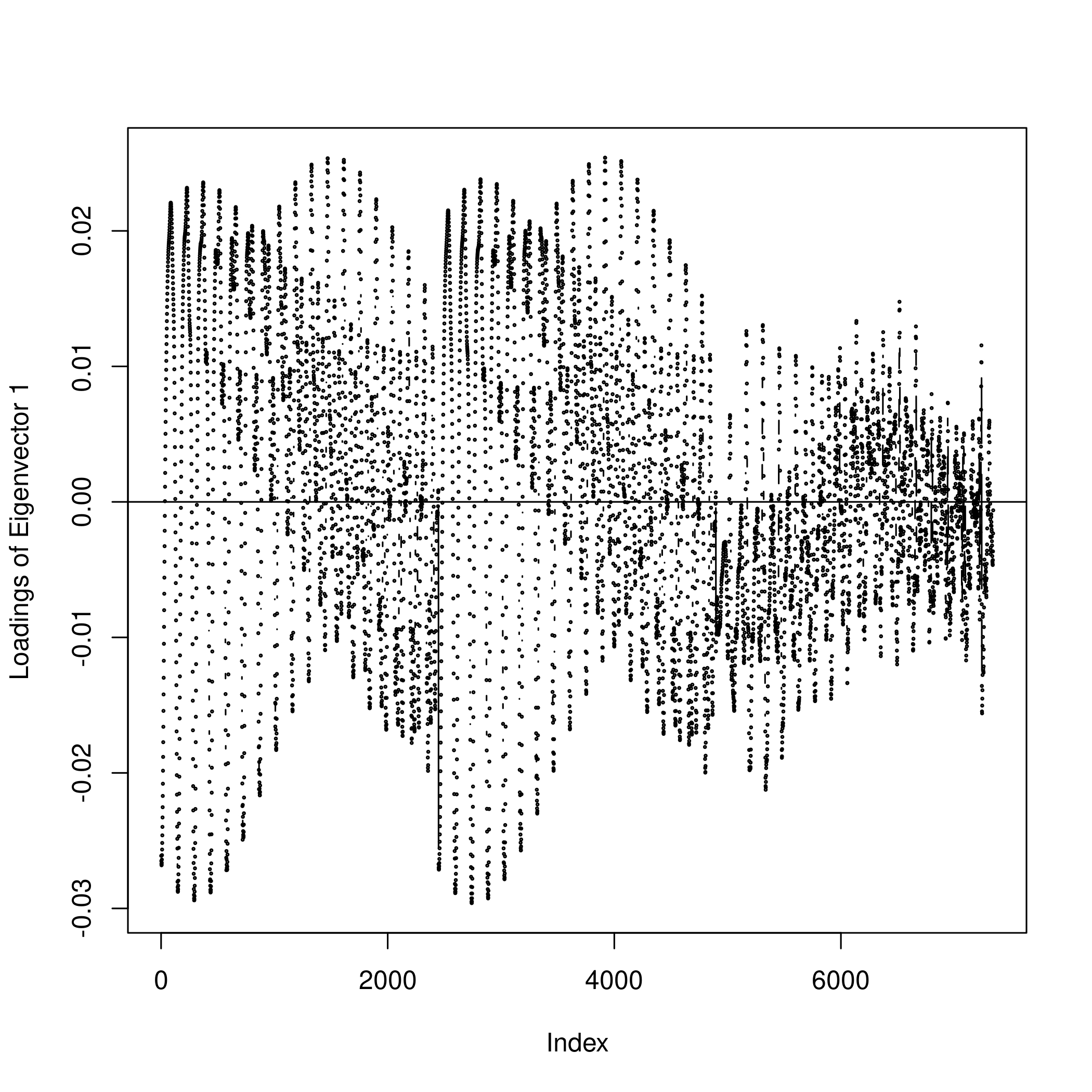} }
Figure 5. The loadings in the first eigenvector, or equivalently, the first eigenface in Figure 4.
The index values (on the x-axis) 1-2448, 2449-4896, and 4897-7344, correspond to $T_v$, $H$,
and $V$, respectively. 

Figure 6 shows these three variances - actually, standard deviations - for all of the eigenvectors.
The black, red, and blue ``curves" correspond to $T_v, H$, and $V$, respectively.  It is important
to note that these variances are {\bf across the loadings}, and not the variances that are
ordinarily associated with each eigenvector (see Figure 3). Said differently, the former are measures 
of {\bf spatial} variability, whereas the latter are measures of temporal variability. It is also 
important to point out that the eigenvectors examined here are eigenvectors of the correlation (not 
covariance) matrix, which means that the analysis is performed on standardized variables; therefore, 
any difference in the three ``curves" in Figure 6 cannot be due to different temporal variances in 
the corresponding fields, but rather due to their spatial correlation structure. 

Consider eigenvector number 1 in Figure 6. 
Consistent with Figure 5, $T_v$ and $H$ have comparable variance across the loadings, both 
larger than the variance of the loadings on $V$. This pattern is true for the first approximately
20 eigenvectors, beyond which the remaining eigenvectors are dominated by $V$. 
 The symmetry about the dashed line is a consequence of the fact that each eigenvector
is normalized to have a magnitude of 1. Said differently, relatively small loadings for some variables
(e.g., $T_v$ and $H$) must be accompanied by relatively large loadings for other variables (e.g., $V$).
Indeed, the dashed line marks the value of a loading in a normalized eigenvector whose loadings are all equal, i.e., $1/\sqrt{L}$, where $L=3 \times 144 \times 17$, is the number of loadings in the 
eigenvector. 

\centerline{ \includegraphics[height=4in,width=4in]{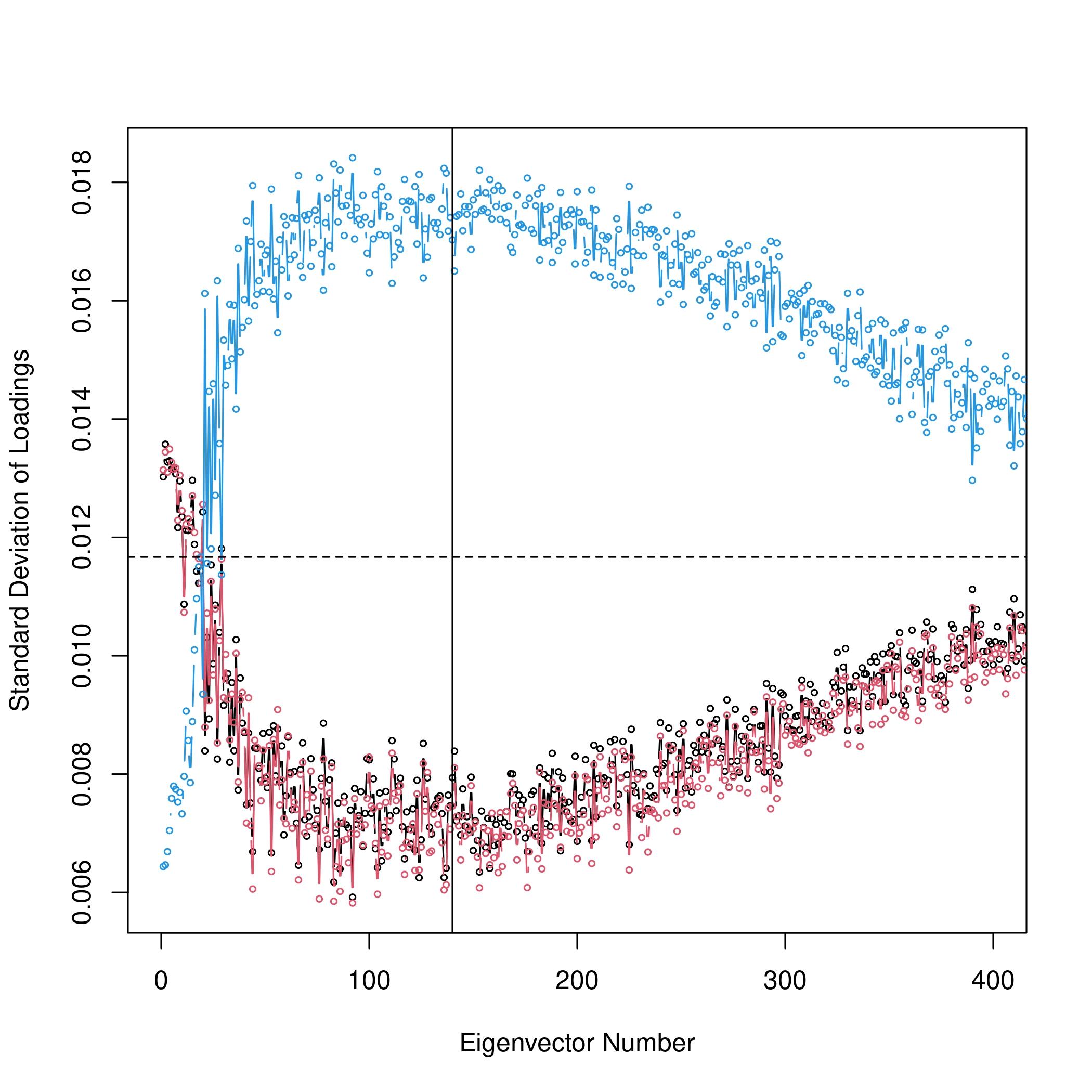} }
Figure 6. The standard deviation of the loadings of $T_v$ (black), $H$ (red), and $V$ (blue),
of each eigenvector. 

As mentioned previously, there is no unique eigenvector that corresponds to hypsometric equation;
one natural criterion for the selection of that eigenvector is suggested by the pattern of 
loadings in Figure 6. Given that Figure 6 displays {\bf spatial} variability, it is reasonable 
to posit that variables relevant to any underlying law must have relatively small spatial variability
because the law is expected to be satisfied at all grid points. This criterion immediately rules
out $V$ as a variable involved in a law because it displays large spatial variability for all
eigenvectors on the right side of the figure. By contrast, $T_v$ and $H$ are ideal candidates for 
a law. Moreover, the
eigenvectors between 100 and 200 have the smallest spatial variability. As such, any one of them
is a potential candidate for the ``last" eigenvector.  

Although, other criteria are proposed in the discussion section, here one can test the validity of 
this criterion because the underlying law is known to be the hypsometric equation (Eq. 5). 
To that end, eigenvector number 150 is selected as the ``last" eigenvector.
Although it is not on the most-right side of the figure, it does have a relatively small eigenvalue 
(denoted with a filled circle in Figure 3).
Using Eq. (3) one can transform a loading ($b$) to a regression coefficient 
($\beta$) estimating the coefficient relating $T_v$ to $H$ in Eq. 5. The histogram of the estimate of 
$\beta$ at all $144 \times 17 = 2,448$ grid points is shown in the top panel of Figure 7. The 
theoretical value, $(R/g) log(850/500) = 15.54 (m/K)$ is also shown as the vertical red line. 
It is evident that the hypsometric law is satisfied at a majority of the grid points. A 2-sided
t-test of the null hypothesis that the mean of the estimated $\beta$ values is equal to the
theoretical value leads to a p-value of 0.93 suggesting that there is no evidence from the
data to justify rejecting the null hypothesis. A 95\% confidence interval for the true mean is
$(13.93, 17.01)$, again consistent with the theoretical value of 15.54. 

The bottom panel in Figure 7 displays the spatial map of the estimates of $\beta$; the lack of a 
coherent spatial structure suggests that the hypsometric equation is satisfied uniformly across the 
spatial field.  All of these results support the proposed criterion for selecting the eigenvector 
corresponding to the hypsometric equation; other criteria are discussed in the next section.

\centerline{ \includegraphics[height=4in,width=4in]{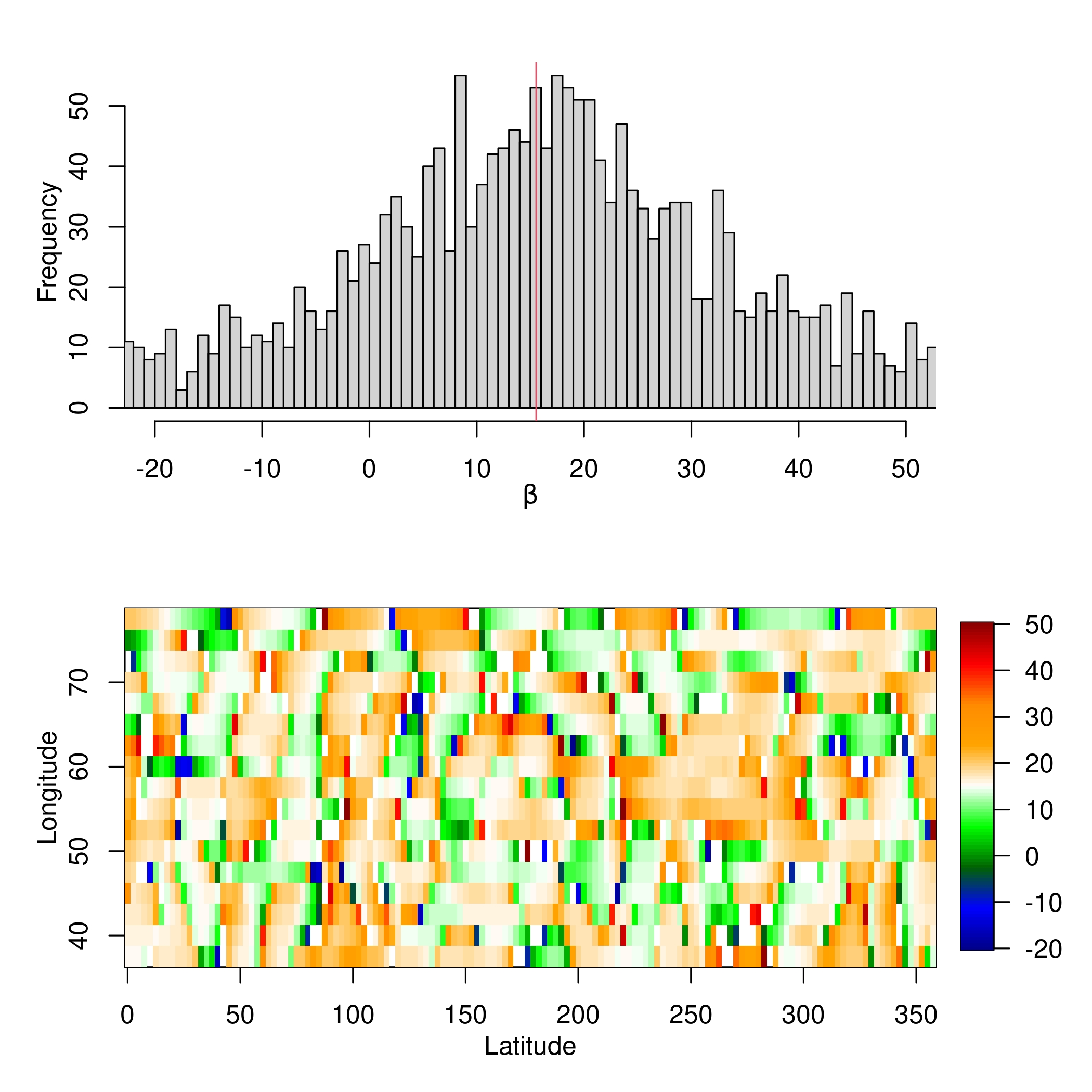} }
Figure 7. The histogram (top) and the spatial map (bottom) of the estimated $\beta$ values in Eq. 3.

\section{Conclusion and Discussion}

It is proposed that PCA can be employed to discover algebraic equations underlying data. Specifically,
it is argued that the eigenvectors with the smallest eigenvalues represent combinations of variables
that are constant across observations, and therefore, can be useful for equation discovery. 
Two examples from Physics and Meteorology are employed to demonstrate the proposal.

One of the limitations of the proposed method is that the underlying equations are assumed to be
linear in the variables (after transforming the variables into log-space). This is a relatively 
broad class of functions, including relations 
of the type $x_0 = \beta_0 + \beta_1 x_1 + \beta_2 x_2 + \cdots $, where $x_i$ denote the variables, 
and $\beta_i$ are the parameters. It also includes relations of the type 
$x_0 = \beta_0 x_1^{\beta_1} x_2^{\beta_2} \cdots$, because such a relationship can be transformed
to the previous one by taking the logarithm of the variables. However, relationships of the
type $x_0 = \beta_0 + \beta_1 x_1^{p_1} + \beta_2 x_2^{p_2} + \cdots$, cannot be discovered by
PCA. Nonlinear extensions of PCA must be considered for that purpose (Hsieh 2009).

The proposed methodology can be extended in a number of ways, specifically in regards to the
criterion for selecting the ``last" eigenvector. In the present work, the criterion is based on the 
spatial variability of the loadings.  Ideally, one may have expected the
last eigenvector to have near-zero loadings on $V$, and large loadings on $T_v$ and $H$. 
That expectation, however, is based on the assumption that the loadings have no spatial variability,
which is simply false. Each of the three fields $T_v, H$ and $V$ have a
nontrivial spatial (within-field) correlation structure that prevents any eigenface to have zero 
variance across grid points.

That said, it is possible to perform the combined PCA on data wherein the $T_v, H$ and $V$ fields
have been decorrelated (or ``whitened"), individually. The result is that the covariance matrix 
{\bf within} each field is diagonal, i.e., no spatial correlation. However, an undesirable 
consequence of this whitening is that the correlations {\bf between} the fields are also affected,
possibly altering the underlying law. To remedy this problem it is possible to formulate the
eigendecomposition as an optimization problem with a penalty term that attempts to
leave the between-field correlations invariant. This idea will be examined at a later time.

Another avenue for future work on the selection of the last eigenvector is based on dimensional 
analysis, more specifically Buckingham's $\pi$-theorem (Bakarji et al. 2022; Bhaskar and Nigam 1990; 
Brence, Dzeroski, and Todorovski 2023; Hardtke 2019; Kasprzak, Lysik, and Rybaczuk 1990; 
Krauss and Wilczek 2014; Xie et al. 2022).
This theorem encapsulates the principle of heterogeneity, according to which the laws of nature can be 
written in terms of dimensionless quantities. Indeed, this theorem can be used to discover the overall 
structure of Kepler's third law and the hypsometric equation, on dimensional grounds alone. 
According to the $\pi$-theorem, it can be shown that the combination $\frac{GMT^2}{a^3}$ in Example 1 
is the only dimensionless quantity that can be constructed from the variables $M, T, a$, and the 
gravitational constant $G$. As such, the only quantity in Kepler's third law that cannot be 
discovered on dimensional grounds is the constant $4\pi^2$. In short, Buckingham's 
$\pi$-theorem can ``explain" why it is possible to discover Kepler's law empirically at all. 
As for the hypsometric equation, first, according to the principle of heterogeneity 
any law (discovered by PCA) must be dimensionless. Second, the $\pi$-theorem implies that the only 
dimensionless quantity that can be constructed from $H, T_v, V$ and the constants $R$ and $g$,
is $\frac{H}{T_v}\frac{g}{R}$; note that the v-component of wind speed $(V)$ does not 
enter that expression. In other words, based on the $\pi$-theorem alone one could have anticipated 
the PCA results above, namely that the variable $V$ is not relevant to the hypsometric equation, and 
also that the quantity relevant to the law is $\frac{H}{T_v}$.  In short, the $\pi$-theorem can aid in 
determining which eigenvector corresponds to a law. Such considerations are especially important
if the data encapsulate multiple laws. Although dimensional analysis has been applied in the
context of regression (Shen et al. 2014; Vignaux and Scott 1999), to the knowledge of the authors
similar work in PCA has not been done. Work is currently in progress to incorporate 
the $\pi$-theorem into PCA in a formal manner. 

In Example 1, Eq. (4) involves a ``normalization" that exposes the near-integer values of the loadings.
That normalization involves multiplying all the elements in a given eigenvector by some constant. In 
that example, it is relatively straightforward to determine the value of that constant for each 
eigenvector. For example, in Eq. (4), all of the elements in the first eigenvector are first divided 
by the first element, and then multiplied by 2. This leads to first and last loadings that are 
integers up to 2 decimal places, and a second loading that is an integer to nearly two decimal places 
(i.e., 2.01); but the third loading (3.83) is relatively far from 4.0. To make this normalization 
process more objective, one can employ Integer Programming (Lenstra 1983; Sierksma and Zwols 2015), 
wherein the parameters of optimization are constrained to be integers. Vines (2000) discusses a 
revision of PCA wherein the loadings are approximately integers. Incorporating these methods into
the proposed method will yield a more effective equation discovery approach.

One issue that has not been addressed here is that of sampling variability. For example, when 3.83
is rounded to the integer 4, is that consistent with the level of noise in the data? In other words, 
it is necessary to provide confidence intervals for the loadings. Then, any integer covered by 
the confidence interval qualifies as a plausible value. Asymptotic results (Anderson 1984; 
Ogasawara 2000) and/or bootstrap methods (Davison, Hinckley, and Schechtman 1986) can be employed
to supplement the proposed method with measures of confidence. 

%

\section{References}

\parindent 0pt

\begin{enumerate}

\item Anderson, T.W., 1963: Asymptotic Theory for Principal Component Analysis.
{\it Ann. Math. Statist.}, {\bf 34 (1)}, 122-148. DOI: 10.1214/aoms/1177704248

\item Anderson, T. W., 1984: {\it An introduction to multivariate statistical analysis} (2nd ed.).
New York: Wiley.

\item Bakarji, J., J. Callaham, S. L. Brunton, and J. N. Kutz: 2022: Dimensionally consistent learning with
Buckingham Pi. {\it Nat. Comput. Sci.}, {\bf 2}, 834–844.

\item Bergen, K. J., P. A. Johnson, M. V. de Hoop, and G. C. Beroza, 2019: Machine learning for
data-driven discovery in solid Earth geoscience. {\it Science}, {\bf 363 (6433)}.
DOI: 10.1126/science.aau0323

\item Bhaskar, R., A. Nigam, 1990: Qualitative physics using dimensional analysis.
{\it Artificial Intelligence}, {\bf 45 (1–2)}, 73-111.

\item Bongard, J., and H. Lipson, 2007: Automated reverse engineering of nonlinear dynamical systems.
{\it Proceedings of the National Academy of Sciences}, {\bf 104 (24)}, 9943-9948.

\item Brence, J., S. Dzeroski, and L. Todorovski, 2023: Dimensionally-consistent equation discovery through
probabilistic attribute grammars. {\it Information Sciences,} {\bf 632}, 742-756.
https://doi.org/10.1016/j.ins.2023.03.073

%

\item Camps-Valls, G., A. Gerhardus, U. Ninad, G. Varando, G. Martius, E. Balaguer-Ballester,
R. Vinuesa, E. Diaz, L. Zanna, J. Runge, 2023: Discovering causal relations and equations from data.
{\it Physics Reports,}, {\bf 1044}, 1-68.

\item Cattell, R. B., 1966: The Scree Test For The Number Of Factors. {\it Multivariate Behavioral
Research}, {\bf 1 (2)}, 245-276. doi:10.1207/s15327906mbr0102\_10.

\item Chen, B., K. Huang, S. Raghupathi, I. Chandratreya, Q. Du, and H. Lipson, 2022:
Automated discovery of fundamental variables hidden in experimental data,
{\it Nature Computational Science}, {\bf 2}, 433-442.

\item Davison, A.C., D.V. Hinkley, and E. Schechtman, 1986: Efficient bootstrap simulation.
{\it Biometrika}, {\bf 73}, 555-566.

\item Dmitrienko, A., C. Chuang-Stein, R. B. D'Agostino, 2007: {\it Pharmaceutical Statistics Using     
SAS: A Practical Guide. SAS Institute}. 380 pp. ISBN 978-1-59994-357-2.

\item Doswell, C. A., III and E. N. Rasmussen, 1994: The Effect of Neglecting the Virtual Temperature
Correction on CAPE Calculations, {\it Wea. Forecasting}, {\bf 9 (4)}, 625-629. https://doi.org/10.1175/1520-0434(1994)009<0625:TEONTV>2.0.CO;2

\item Elmore, K. L., and M. B. Richman, 2001: Euclidean Distance as a Similarity Metric for
Principal Component Analysis. {\it Mon. Wea. Rev.}, {\bf 129 (3)}, 540-549.
https://doi.org/10.1175/1520-0493(2001)129<0540:EDAASM>2.0.CO;2

\item Gertler, J., and J. Cao, 2005: Design of optimal structured residuals from partial principal
component models for fault diagnosis in linear systems. {\it Journal of Process Control},
{\bf 15 (5)}, 585-603.

\item Glickman, T. 2000: Glossary of meteorology. American Meteorological Society. ISBN: 9781878220349.
https://glossary.ametsoc.org/wiki/Virtual\_temperature\#:~:text=Hence\%20the\%20virtual\%20temperature\%20is,\%2B

\item Grundner, A., T. Beucler, P Gentine, V. Eyring, 2023: Data-Driven Equation Discovery of a Cloud Cover Parameterization, arXiv preprint arXiv:2304.08063

\item Hardtke, J-D., 2019: On Buckingham’s $\pi$-Theorem.  http://arxiv.org/abs/1912.08744v1

\item Hoerl, R. W., R. D. Snee, R. D. De Veaux, 2014: Applying statistical thinking to ‘Big Data’ problems.
{\it WIREs Computational Statistics}, {\bf 6 (4)}, 211-312.
https://doi-org.offcampus.lib.washington.edu/10.1002/wics.1306open\_in\_new

\item Hotelling, H. (1933). Analysis of a complex of statistical variables into principal components. Journal of Educational Psychology, 24(6), 417–441. https://doi.org/10.1037/h0071325

\item Hsieh, W.W., 2009: Nonlinear Principal Component Analysis. In: Haupt, S.E., Pasini, A., Marzban, C.
(eds) {\it Artificial Intelligence Methods in the Environmental Sciences}. Springer, Dordrecht. https://doi.org/10.1007/978-1-4020-9119-3\_8

\item Huang, B., 2001: Process identification based on last principal component analysis. {\it Journal of
Process Control}, {\bf 11}, 19-33.

\item Ibebuchi, C. C., M. B. Richman, 2023: Circulation typing with fuzzy rotated T-mode principal
component analysis: methodological considerations. {\it Theoretical and Applied Climatology},
{\bf 153}, 495-523.

\item Jolliffe I. T. 2002: {\it Principal Component Analysis}, 2nd edn. New York, NY: Springer-Verlag. 518 pp.

\item Jolliffe I. T., J. Cadima, 2016: Principal component analysis: A review and recent developments.
{\it Phil. Trans. R. Soc. A}, {\bf 374}. Article ID: 20150202. http://dx.doi.org/10.1098/rsta.2015.020
site Jolliffe's book on pca, and their own section 3c for an image analysis example.

\item Kalnay E., M. Kanamitsu, R. Kistler, W. Collins, D. Deaven, L. Gandin, M. Iredell, S. Saha, G. White,
J. Woollen, Y. Zhu, M. Chelliah, W. Ebisuzaki, W. Higgins, J. Janowiak, K. C. Mo, C. Ropelewski,
J. Wang, A. Leetmaa, R. Reynolds, Roy Jenne, and Dennis Joseph, 1996: The NCEP/NCAR 40-Year Reanalysis
Project. {\it BAMS}, {\bf 77(3)}, 437–472.
https://doi.org/10.1175/1520-0477(1996)077<0437:TNYRP>2.0.CO;2

\item Kamath, C., 2001: On Mining Scientific Datasets. In: Grossman, R.L., Kamath, C., Kegelmeyer, P.,
Kumar, V., Namburu, R.R. (eds) {\it Data Mining for Scientific and Engineering Applications}. Massive Computing, vol 2. Springer, Boston, MA. https://doi.org/10.1007/978-1-4615-1733-7\_1

\item Kasprzak, W., B. Lysik, and M. Rybaczuk, 1990: {\it Dimensional Analysis in the
Identification of Mathematical Models,}  204 pp.  https://doi.org/10.1142/1162

\item Kratzert, F., D. Klotz, G. Shalev, G. Klambauer, S. Hochreiter, and G. Nearing, 2019:
Towards learning universal, regional, and local hydrological behaviors via machine learning applied
to large-sample datasets. {\it HESS}, {\bf 23 (12)}, 5089–5110. https://doi.org/10.5194/hess-23-5089-2019

\item Krauss, L.M. and F. Wilczek, 2014: Using cosmology to establish the quantization of            
gravity.  {\it Phys. Rev. D}, {\bf 89}, 047501.

\item Kurgan, L., and P. Musilek, 2006: A survey of Knowledge Discovery and Data Mining process models. 
{\it The Knowledge Engineering Review}, {\bf 21 (1)}, 1–24. 

\item Langley, P. 1981: Data-driven discovery of physical laws.
{\it Cognitive Science}, {\bf 5 (1)}, 31-54. https://doi.org/10.1016/S0364-0213(81)80025-0

\item Lenstra, H. W., 1983: Integer Programming with a Fixed Number of Variables. {\it Mathematics of
Operations Research}, {\bf 8 (4)}, 538-548.  doi:10.1287/moor.8.4.538.

\item Lorenz, E. N. 1956. Empirical orthogonal functions and statistical weather prediction, Scientific
Report 1, Statistical Forecasting Project. Massachusetts Institute of Technology Defense Doc. Center
No. 110268, 49 pp.  http://muenchow.cms.udel.edu/classes/MAST811/Lorenz1956.pdf .

\item Marzban, C., and U. Yurtsever, 2017: On the Shape of Data. Paper presented at the 15th Conference on Artificial Intelligence, at the 97th American Meteorological Society Annual Meeting, Seattle, Jan. 22-26.

\item Marzban, C., and U. Yurtsever 2011: Baby Morse theory in data analysis. Paper at the workshop on Knowledge Discovery, Modeling and Simulation (KDMS), held in conjunction with the 17th ACM SIGKDD Conference on Knowledge Discovery and Data Mining, San Diego, CA., August 21-24.

\item North, G. R., T. L. Bell, R. F. Cahalan, and F. J. Moeng, 1982: Sampling Errors in the Estimation of
Empirical Orthogonal Functions. {\it Mon. Wea. Rev.}, {\bf 110 (7)}, 699-706.
DOI: https://doi.org/10.1175/1520-0493(1982)110<0699:SEITEO>2.0.CO;2

\item Ogasawara, H., 2000: Standard errors of the principal component loadings for unstandardized and
standardized variables. {\it British Journal of Mathematical and Statistical Psychology}, {\bf 53},
155-174.

\item Rolle, J.-D., 2002: Notes about the last principal component. {\it Appl. Math. Comput.}, {\bf 126}, 231-241.

\item Schmidt, M., and H. Lipson, 2009: Distilling Free-Form Natural Laws from Experimental Data.
{\it Science}, {\bf 324}, 81-85.

\item Shen, W., T. Davis, D. K. J. Lin, and C. J. Nachtsheim, 2014: Dimensional Analysis and Its
Applications in Statistics. {\it Journal of Quality Technology}, {\bf 46 (3)}, 185-198.

\item Sierksma, G, and Y. Zwols, 2015: {\it Linear and Integer Optimization: Theory and Practice.}
CRC Press. ISBN 978-1-498-71016-9.

\item Song W., L. Shi, X. Hu, Y. Wang, and L. Wang, 2023: Reconstructing the Unsaturated Flow Equation
From Sparse and Noisy Data: Leveraging the Synergy of Group Sparsity and Physics-Informed Deep Learning.
{\it Water Resources Research}, {\bf 59(5)}, 1-24. https://doi.org/10.1029/2022WR034122

\item Vignaux, G. A., and J. L. Scott, 1999: Simplifying Regression Models Using Dimensional Analysis. 
{\it The Australian and New Zealand Journal of Statistics}, {\bf 41 (1)}, 31–42.

\item Vines, S.K., 2000: Simple principal components. {\it Appl. Statist.},  {\bf 49 (4)}, 441-451.

\item Wallace, J. M., and P. V. Hobbs, 1977: {\it Atmospheric Science: An Introductory Survey}.
Academic Press.

\item Wang, X.Z., 1999: Data mining and knowledge discovery for process monitoring and control. 
{\it Advances in Industrial Control}, 1-251, ISBN 1-85233-137-2 .

\item Wilks, D. S., 2016: Modified ``Rule N" Procedure for Principal Component (EOF) Truncation.
{\it Journal of Climate}, {\bf 29 (8)}, 049–3056. https://doi.org/10.1175/JCLI-D-15-0812.1

\item Wilks, D. S., 2019: {\it Statistical Methods in the Atmospheric Sciences}, Fourth Edition, Elsevier.
https://doi.org/10.1016/C2017-0-03921-6

\item Xie X., A. Samaei, J. Guo, W. K. Liu, and Z. Gan, 2022: Data-driven discovery of dimensionless
numbers and governing laws from scarce measurements. {\it Nature Communications}, {\bf 13}, 7562.

\item Xu, W., and M. Stalzer, 2019: Deriving compact laws based on algebraic formulation of a
data set.  {\it Journal of Computational Science}, {\bf 36}. https://doi.org/10.1016/j.jocs.2019.06.006

\item Yu, S., and J. Ma, 2021: Deep learning for geophysics: Current and future trends. {\it Reviews of
Geophysics}, {\bf 59}. https://doi.org/10.1029/2021RG000742

\item Zanna, L., and T. Bolton, 2020: Data‐driven equation discovery of ocean mesoscale closures. 
{\it Geophys. Res. Lett.}, {\bf 47}, e2020GL088376, https://doi.org/10.1029/2020GL088376.

\item Zhang, S., and G. Lin, 2018: Robust data-driven discovery of governing physical laws with error bars.
{\it Proceedings of the Royal Society A}, {\bf 474}.  https://doi.org/10.1098/rspa.2018.0305

\item Zlobina, A.G., and I. V. Zhurbin, 2020: Applying a Principle Component Analysis to Search for Objects on Historical Territories by the Spectral Brightness of Vegetation. {\it J. Phys.: Conf. Ser.},
{\bf 1611}, 1-7. doi:10.1088/1742-6596/1611/1/012064

\end{enumerate}
\end{document}